\documentclass[a4paper,11pt]{article}
\usepackage{pos}

\def\ubar{\overline{u}}

\title{Null test BSM searches with rare charm baryon decays}

\author*{Marcel Golz}
\author{Gudrun Hiller}
\author{Tom Magorsch}

\affiliation{TU Dortmund University, Department of Physics, Otto-Hahn-Str.4, D-44221 Dortmund, Germany}

\emailAdd{marcel.golz@tu-dortmund.de}
\emailAdd{ghiller@physik.uni-dortmund.de}
\emailAdd{tom.magorsch@tu-dortmund.de}

\abstract{Rare $\vert \Delta c \vert=\vert \Delta u \vert=1$ processes  uniquely probe  flavor in the Standard Model and beyond from the up-type quark sector.
Opportunities to search for BSM physics in charm arise from the severe GIM suppression which kills SM contributions to leptonic axial-vector contributions
and suppresses CP violation.
Semileptonic decays of charmed hadrons offer a variety of clean null test observables,
featuring new physics effects which are even enhanced by resonance contributions. In particular, angular observables in three- and self-analyzing four-body baryon decays, such as $\Lambda_c\to p \ell^+\ell^-$ and $\Xi_c^+\to\Sigma^+(\to p\pi^0)\ell^+\ell^-$, $\Xi_c^0\to\Lambda^0\,(\to p \pi^-)\ell^+\ell^-$ and  
$\Omega_c^0\to\Xi^0\,(\to \Lambda^{0} \pi^0)\ell^+\ell^-$ disentangle possible new physics effects in electromagnetic dipole couplings $C_7^{(\prime)}$ and (axial-)vector 4-fermion ones $C^{(\prime)}_{9\,(10)}$. There is sensitivity to BSM couplings as small as  $\sim 0.01$.
}

\FullConference{%
  11th International Workshop on the CKM Unitarity Triangle (CKM2021)\\
  22-26 November 2021\\
  The University of Melbourne, Australia
}

\begin{document}
\rightline{DO-TH 22/08}
\maketitle

\section{Introduction}
\noindent Semileptonic rare charm decays are strongly Glashow-Iliopoulos-Maiani (GIM) suppressed in the Standard Model (SM) due to light down-type quarks running in one-loop diagrams~\cite{Burdman:2001tf}. Therefore, Beyond Standard Model (BSM) effects can be probed in $c\to u\ell^+\ell^-$ modes. However, several impediments need to be overcome in BSM searches in rare charm decays. These include the lack of sufficient theoretical control on decay amplitudes due to strong pollution from intermediate QCD resonances and poor convergence of the heavy quark expansion at the charm mass scale~\cite{Feldmann:2017izn}. Null test observables, already part of  the ongoing precision programs of $b\to s\ell^+\ell^-$ transitions, vanish in the SM and hence any signal indicates BSM physics, which makes null tests inevitable in New Physics (NP) searches in $\vert \Delta c \vert=\vert \Delta u \vert=1$ processes.
In particular, the spin structure in rare charm baryon decays offers a rich angular distribution, including several clean null tests~\cite{Golz:2021imq, Golz:2022alh}.
Recent experimental and theoretical progress of rare charm decays is compiled in Ref.~\cite{Gisbert:2020vjx}. Research interest in the field is rapidly increasing, see \textit{e.g.}~\cite{Burdman:2001tf, Gisbert:2020vjx, deBoer:2015boa, Fajfer:2015mia, Feldmann:2017izn, Bause:2019vpr, Bharucha:2020eup, Golz:2021imq, Golz:2022alh}.

The plan of this paper is as follows. The effective field theory framework and SM phenomenology in rare charm baryon decays is presented in Sec.~\ref{sec:eft_pheno}. Sec.~\ref{sec:angular} discusses the NP sensitivity from null tests in angular observables of $\Lambda_c\to p \mu^+\mu^-$. We discuss further opportunities in rare charm four-body decay modes based on recent results in Ref.~\cite{Golz:2022alh} and conclude in Sec.~\ref{sec:concl}.

\section{Standard Model phenomenology of rare charm baryon decays}\label{sec:eft_pheno}
\noindent The effective Hamiltonian at the charm mass scale $\mu_c$ inducing $c\to u\mu^+\mu^-$ processes reads
\begin{equation}
\mathcal{H}_{\rm eff} \supset -\frac{4G_F}{\sqrt2} \frac{\alpha_e}{4\pi} \sum_{k=7,9,10} \bigl( C_kO_k + C_k^\prime O_k^\prime \bigr)\,,
\label{eq:Heff}
\end{equation}
where the dimension 6 operators relevant for this work are defined as follows:
\begin{equation}
\begin{split}
O_7 &= \frac{m_c}{e} (\ubar_L \sigma_{\alpha\beta} c_R) F^{\alpha\beta} \,,\\
O_9 &= (\ubar_L \gamma_\mu c_L) (\overline{\mu} \gamma^\mu \mu) \,,\\
O_{10} &= (\ubar_L \gamma_\mu c_L) (\overline{\mu} \gamma^\mu \gamma_5 \mu) \,, \\
\end{split}
\quad\quad
\begin{split}
O^\prime_7 &= \frac{m_c}{e} (\ubar_R \sigma_{\alpha\beta} c_L) F^{\alpha\beta}\,,  \\
O^\prime_9 &= (\ubar_R \gamma_\mu c_R) (\overline{\mu} \gamma^\mu \mu) \,, \\ 
O^\prime_{10} &= (\ubar_R \gamma_\mu c_R) (\overline{\mu} \gamma^\mu \gamma_5 \mu) \,, \\
\end{split}
\label{eq:operators}
\end{equation}
with $\sigma_{\alpha\beta}=\frac{\text{i}}{2}\left[\gamma_\alpha,\,\gamma_\beta\right]$ and the electromagnetic field strength tensor $F^{\alpha\beta}$.

Short distance contributions to operators in Eq.~\eqref{eq:operators} in the SM are calculated consistently up to (partly) NNLLO in RG improved perturbation theory in Ref.~\cite{deBoer:thesis}.  All contributions originate solely from four-quark operators at the $W$ mass scale and are contained in effective $q^2$ dependent coefficients $C_7^{\text{eff}}(q^2),\,C_9^{\text{eff}}(q^2)$, where $q^2$ is the dimuon invariant mass squared and $C_7^{\text{eff}}\lesssim 10^{-3}$ and $C_9^{\text{eff}}(q^2>0.1\,\text{GeV}^2)\lesssim 10^{-2}$, negligible for most phenomenological applications, see results in Ref.~\cite{deBoer:thesis} and discussions in~\cite{Golz:2021imq}. SM contributions to all primed operators and $C_{10}$ vanish. 

Further contributions to $\vert \Delta c\vert=\vert\Delta u \vert = 1 $ come from intermediate resonances, which we parametrize in terms of a sum of Breit-Wigner distributions fit to data. These resonance effects are then added as a contribution to $O_9$ in $C_9^R(q^2)$ and are the main source of uncertainty due to unknown strong phases entering the parametrization. Dominant vector resonances include $M=\rho(770)$, $\omega(782)$ and $\phi(1020)$ and parameters are fixed via $\mathcal{B}_{C_9^R}(\Lambda_{c} \to  p\mu^{+}\mu^{-}) =  \mathcal{B}(\Lambda_{c} \to pM)\mathcal{B}(M \to \mu^{+}\mu^{-})\,$. Contributions from pseudoscalar mesons $\eta,\,\eta^\prime$ are found to be negligible on the level of branching ratios in rare charm baryon decays~\cite{Golz:2021imq}.

\begin{figure}[!t]\centering
\includegraphics[width=0.45\textwidth]{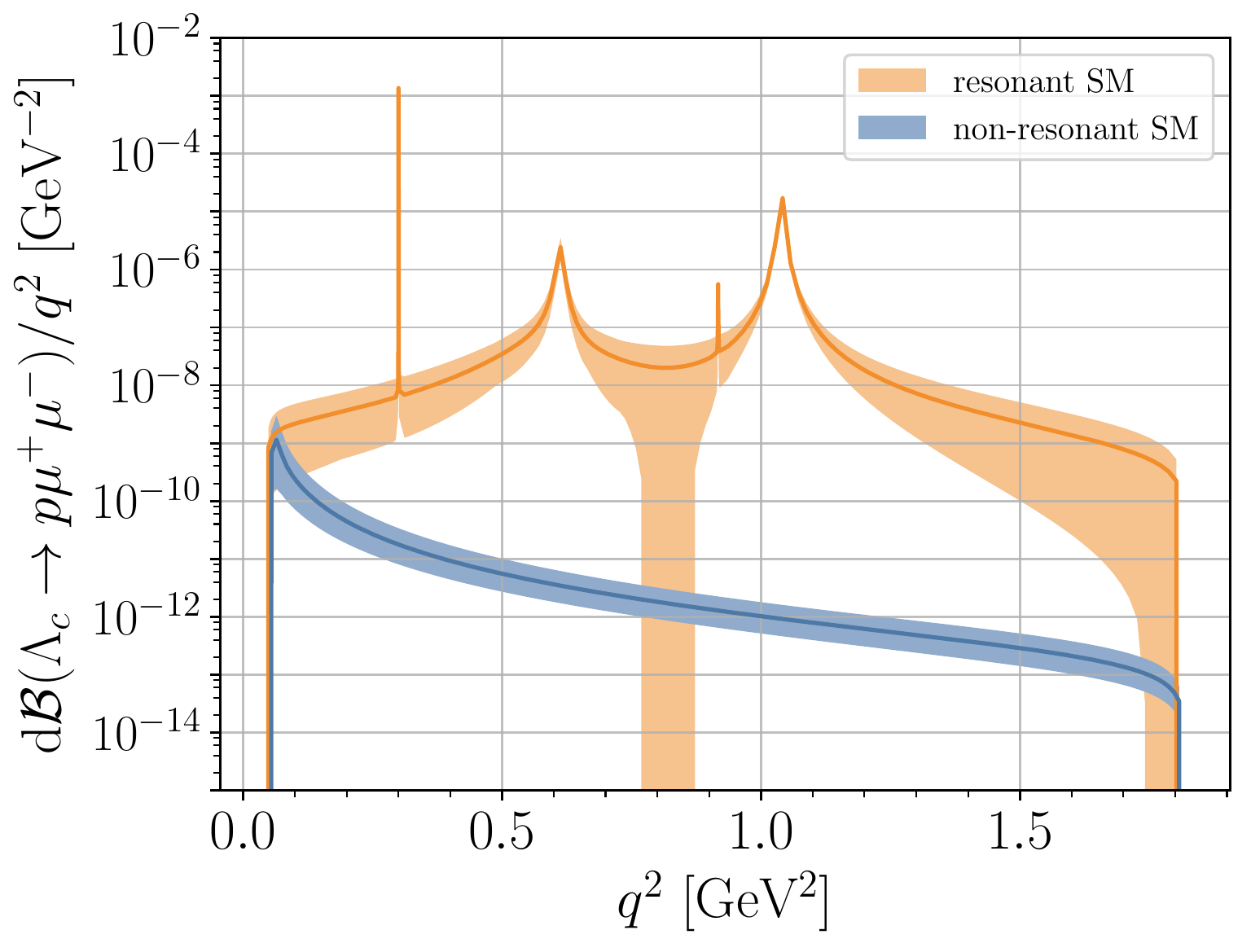}
\caption{The resonant (orange) and non-resonant (blue)  $q^2$-differential branching ratio in the  SM for the rare charm baryon decay $\Lambda_c\to p \mu^+\mu^-$. The bandwidths indicate uncertainties from form factors and strong phase variation (charm mass scale $\mu_c$ variation) in the (non-)resonant case. Figure taken from Ref.~\cite{Golz:2021imq}.}
\label{fig:br}
\end{figure}

We demonstrate the dominance of long-range resonances over non-resonant short-distance SM contributions in Fig.~\ref{fig:br}, where the $q^2$ differential branching ratio for $\Lambda_c \to p \mu^+\mu^-$ is shown in orange and blue, respectively.
The main sources of uncertainty are from form factors~\cite{Meinel:2017ggx} and due to varying strong phases independently from $-\pi$ to $\pi$ for resonant contributions and due to varying the charm mass scale $\mu_c$ for short-distance SM effects.

\section{Null test observables in angular distributions and beyond}\label{sec:angular}

\noindent The angular distribution of three-body rare charm baryon decays can be described by the dilepton invariant mass $q^2$ and a single angle $\theta_\ell$ associated to the angle between the $\ell^+$ momentum and the negative direction of flight of the $\Lambda_c$ in the dilepton rest frame. Then, contributions are split into angular coefficients as
\begin{equation}
  \frac{\text{d}^2\Gamma}{\text{d}q^2\text{d}\cos\theta_\ell}=\frac{3}{2}\,(K_{1ss}\,\sin^2\theta_\ell\,+\,K_{1cc}\,\cos^2\theta_\ell\,+\,K_{1c}\,\cos\theta_\ell)\,
  \label{eq:angl_distr}
\end{equation}
where the dependence on Wilson coefficients and form factors of $K_{1ss},\,K_{1cc}$ and $K_{1c}$ is given in Ref.~\cite{Golz:2021imq}. Next to the differential branching ratio, two angular observables can be defined, the fraction of longitudinally polarized dimuons $F_L$ and the forward-backward asymmetry in the leptons $A_{\text{FB}}$, defined as
\begin{align}\label{eq:angulars}
F_L=\frac{2\,K_{1ss}-K_{1cc}}{2\,K_{1ss}+K_{1cc}}\,, \quad A_{\text{FB}}=\frac{3}{2}\,\frac{K_{1c}}{2\,K_{1ss}+K_{1cc}}\,,
\end{align} 
where $\frac{\text{d}\Gamma}{\text{d}q^2}=2\,K_{1ss}+K_{1cc}$, and one can also define $\tilde{A}_{\text{FB}}=\frac{3}{2}\frac{K_{1c}}{\Gamma}$ with $\Gamma=\int_{q_{\text{min}}^2}^{q_{\text{max}}^2}(2\,K_{1ss}+K_{1cc})\text{d}q^2$.

\begin{figure}[!t]\centering
\includegraphics[width=0.42\textwidth]{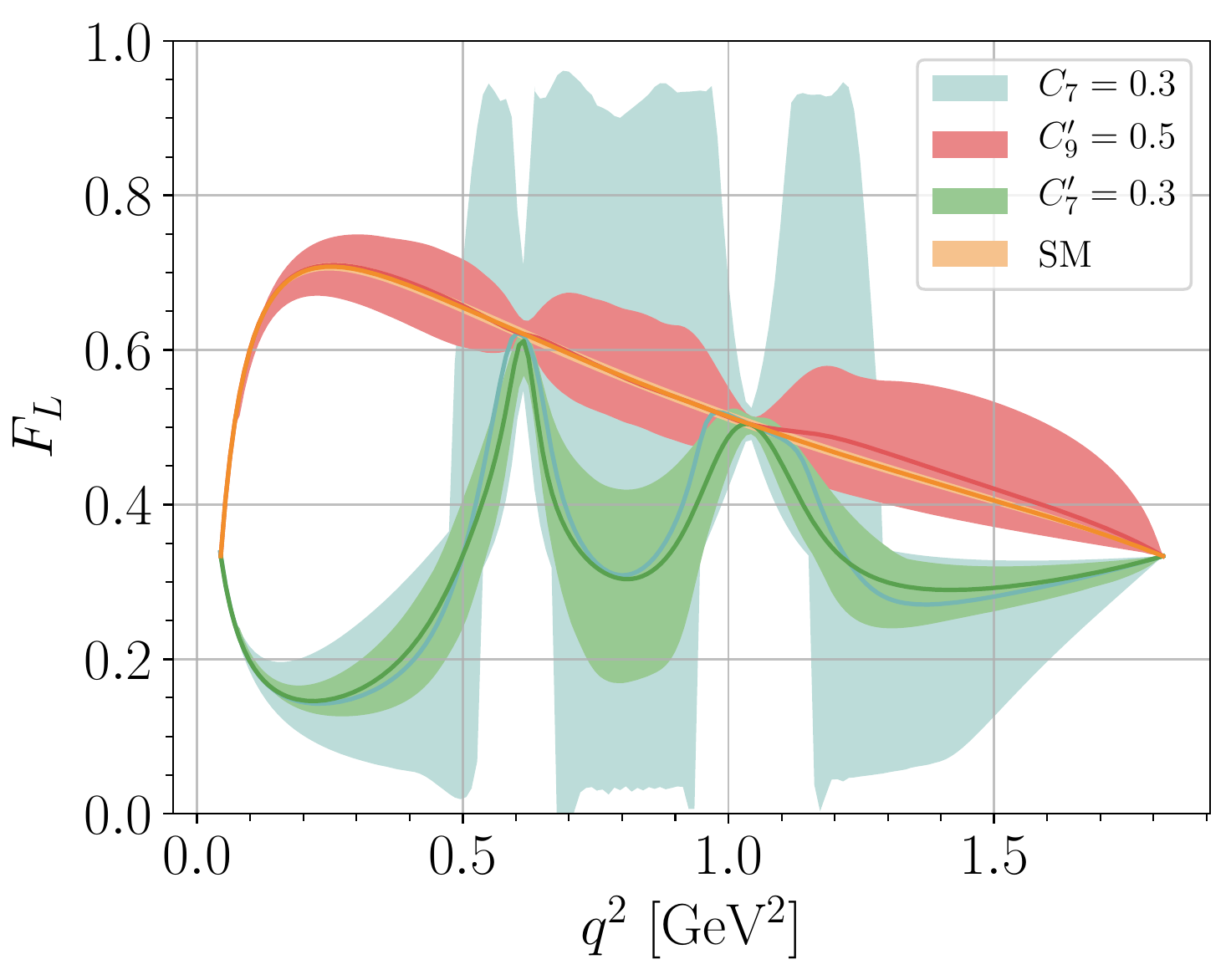}
\includegraphics[width=0.42\textwidth]{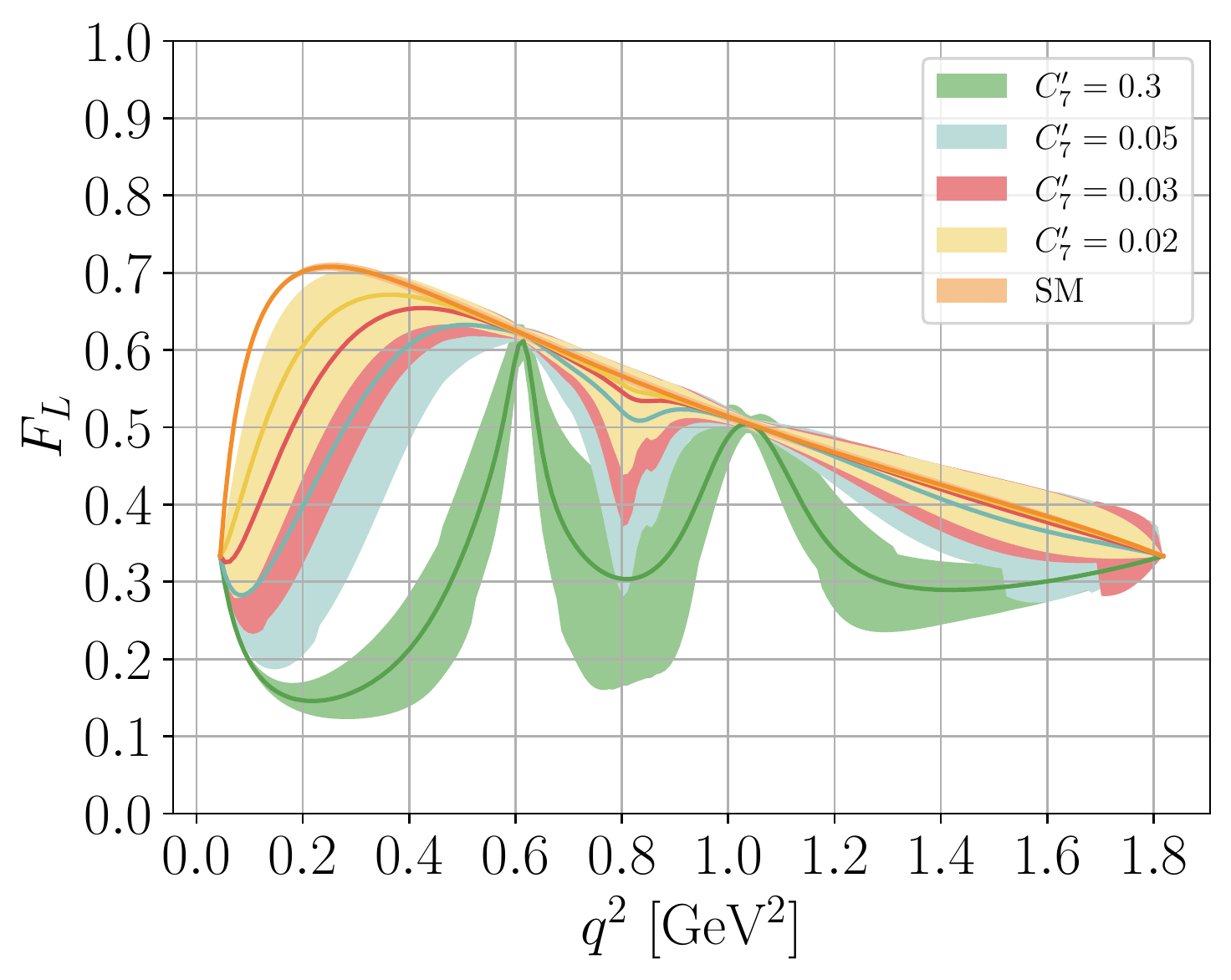}
\includegraphics[width=0.42\textwidth]{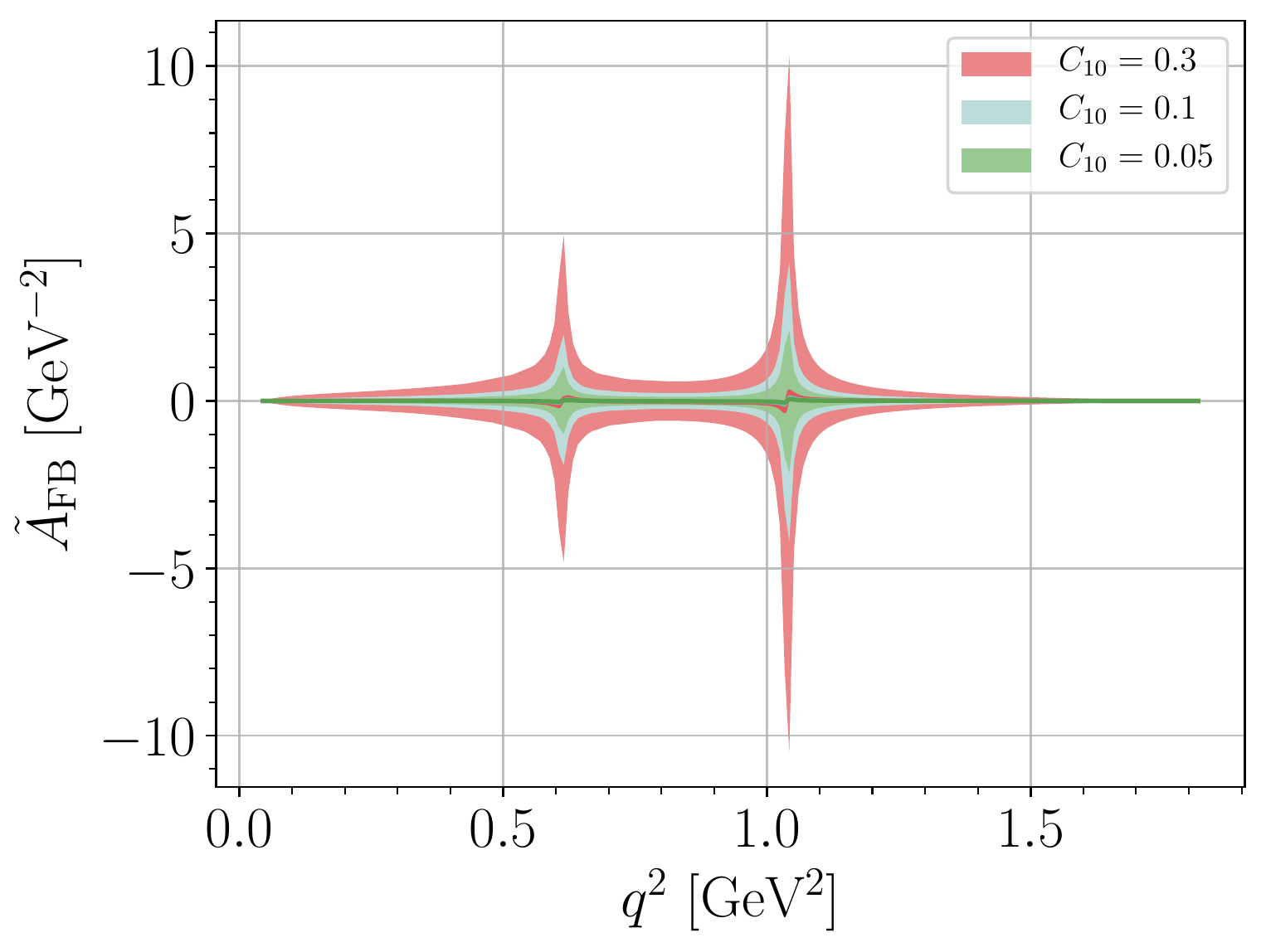}
\includegraphics[width=0.42\textwidth]{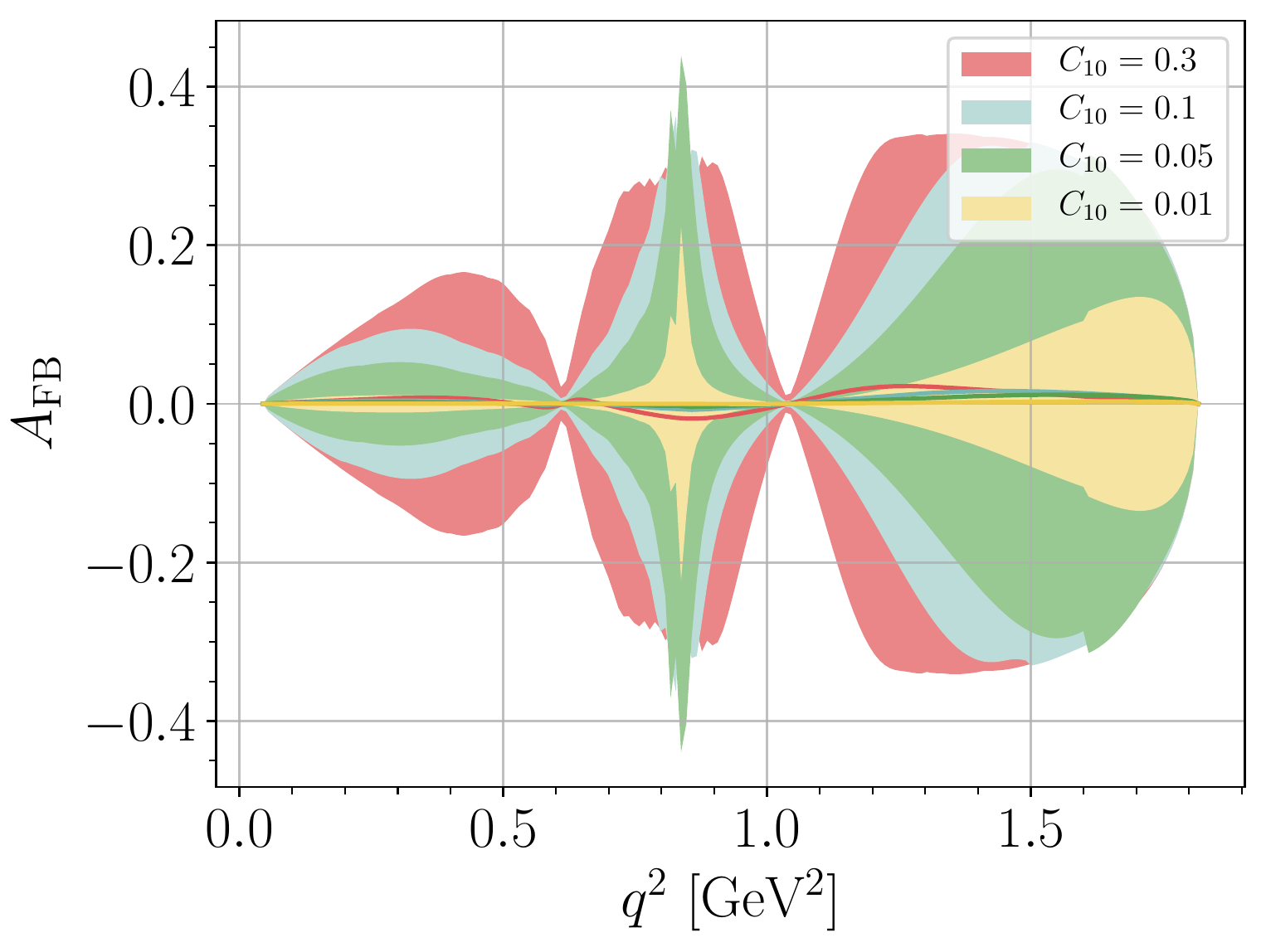}
\caption{Angular observables in $\Lambda_c\to p\mu^+\mu^-$. The upper row shows the fraction of longitudinally polarized dimuons $F_L$ in various NP scenarios (left) and the sensitivity for decreasing values of the dipole operator $C_7^\prime$. The lower row shows the forward-backward asymmetry $\tilde{A}_{\text{FB}}$ (left) and ${A}_{\text{FB}}$ (right) in NP scenarios with decreasing $C_{10}$ Wilson coefficient, see text for details. Upper left and lower row plots are taken from Ref.~\cite{Golz:2021imq}.}
\label{fig:ang}
\end{figure}

We show $F_L$ (top row), $\tilde{A}_{\text{FB}}$ (bottom left) and ${A}_{\text{FB}}$ (bottom right) in the SM and in several BSM benchmark scenarios in Fig.~\ref{fig:ang}.

For $F_L$, we learn that uncertainties associated with resonances drop out in the SM and $F_L=\frac{1}{3}$ model-independently at both kinematic endpoints~\cite{Hiller:2021zth,Golz:2021imq}. The $q^2$-shape is significantly altered from the one in the SM  in scenarios involving dipole couplings $C_7^{(\prime)}$. There is  sensitivity to values as low as $\sim 0.02$, as illustrated in the upper right plot. $A_{\text{FB}}$ is a SM null test, which is why no orange SM curve is shown (it is exactly zero). Depending on the normalization, the axial vector coupling $C_{10}$ can also be probed down to $\sim 0.01$. In $\tilde{A}_{\text{FB}}$, the normalization is a constant in $q^2$, and the signal is resonance enhanced, due to large values of $C_9^R$ around the resonance masses interfering with the BSM coupling $C_{10}$. This resonance enhancement can not be seen in $A_{\text{FB}}$ with differential normalization as the peaking contributions in the differential branching ratio around the resonances suppress the signal in the numerator.

\section{Opportunities with self-analyzing four-body decays}\label{sec:fourbody}

\begin{figure}[!t]\centering
\includegraphics[width=0.41\textwidth]{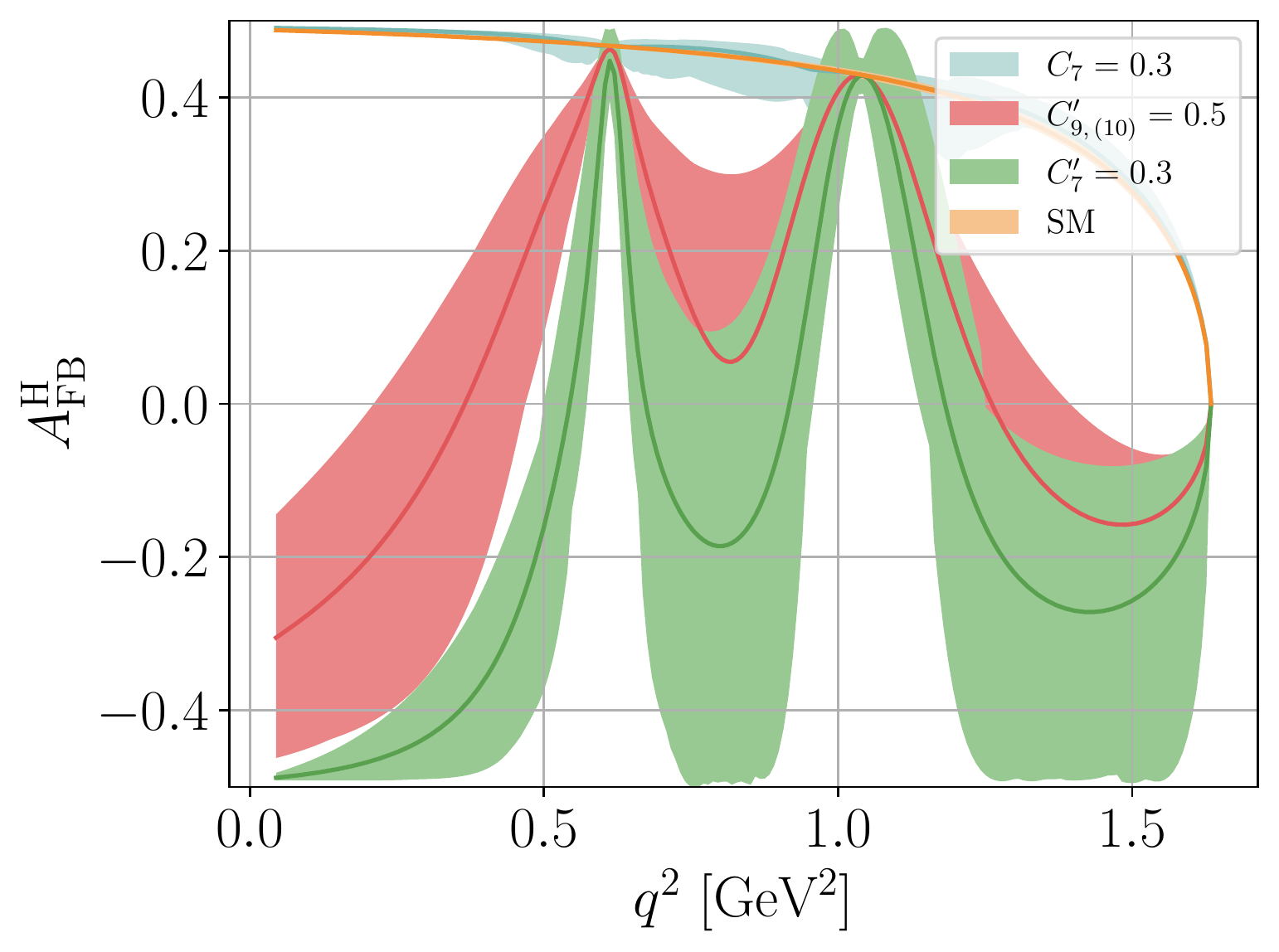}\\
\includegraphics[width=0.41\textwidth]{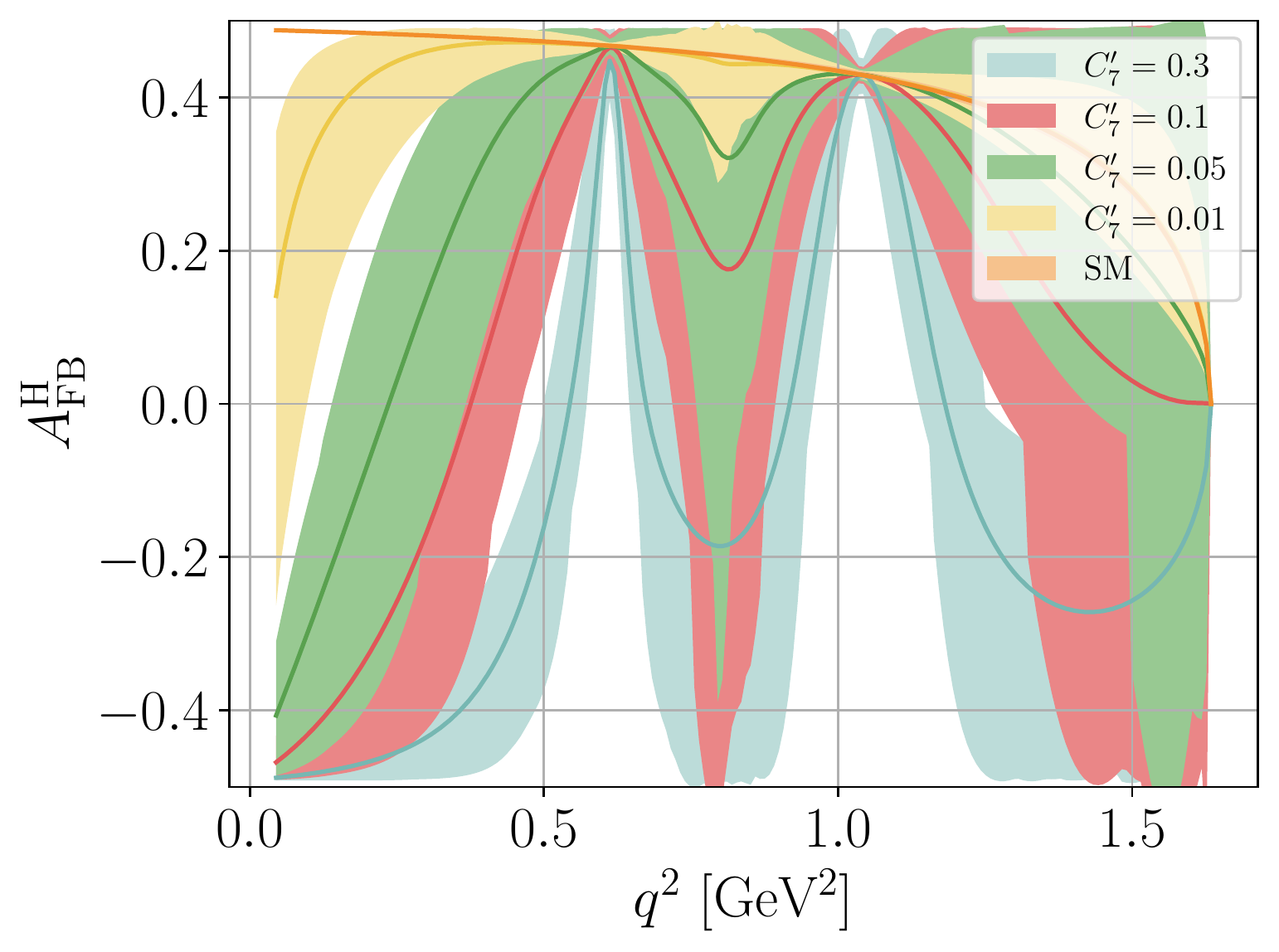}
\includegraphics[width=0.41\textwidth]{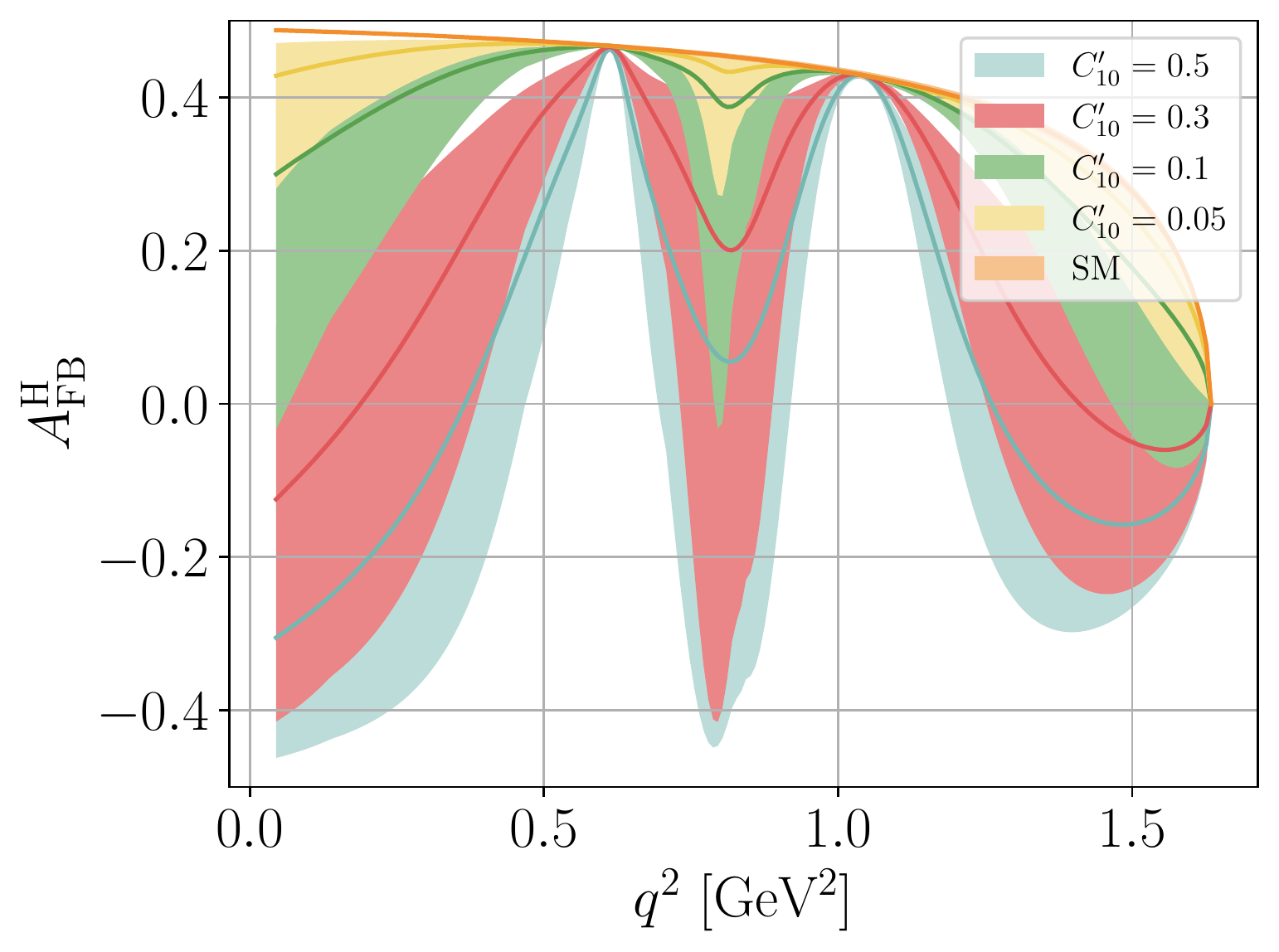}
\caption{$A_{\text{FB}}^{\text{H}}$ for $\Xi_c^+\to \Sigma^+(\to p\pi^0) \mu^+\mu^-$. The upper plot is taken from Ref.~\cite{Golz:2022alh} and shows several BSM scenarios (see legend) and the SM. The lower row plots further illustrate the sensitivity to $C_7^\prime$ and $C_{9,\,(10)}^\prime$ via decreasing benchmark values. The SM, shown in orange, includes form factor uncertainties, whereas uncertainties from resonance parameters cancel. In NP scenarios, the dominant uncertainties are due to the strong phases associated to $C_9^R$ and entering via interference with the BSM Wilson coefficients.}
\label{fig:ang2}
\end{figure}

\noindent Recently, further null test opportunities were presented for (quasi-) four-body decay chains of charmed baryons, where the final state baryon further decays weakly with sizable polarization parameter $\alpha$~\cite{Golz:2022alh}. These \textit{self analyzing} four-body final states offer a rich angular distribution with complementary sensitivities to NP couplings. We refer to Ref.~\cite{Golz:2022alh} for further details, but exemplary discuss for the decay chain $\Xi_c^+\to \Sigma^+(\to p\pi^0) \mu^+\mu^-$ the forward-backward asymmetry of the final state hadron system $A_{\text{FB}}^{\text{H}}$, which can be defined in close analogy to $A_{\text{FB}}$ in Eq.~\eqref{eq:angulars}. $A_{\text{FB}}^{\text{H}}$ is not a null test, but very sensitive to right-handed quark currents $C_{7,\,9,\,10}^\prime$. This can be seen in Fig.~\ref{fig:ang2}, where the upper plot shows the SM in orange and contributions for sizable right-handed currents in red and green, compared to small deviations induced by NP in $C_7$ in blue, whereas NP benchmarks with $C_9$ or $C_{10}$ are not shown, as they are SM--like. The lower plots again illustrate sensitivities, the plot to the left for $C_7^\prime$ and the plot to the right for $C_{10}^\prime$, but a plot for $C_9^\prime$ would be equal to the latter within uncertainties. As evident from the lower row plots, $C_{7}^\prime$ can be probed down to $\sim 0.01$, whereas the sensitivity in $C_{9,\,(10)}^\prime$ is $\mathcal{O}(0.1)$.

Beyond $A_{\text{FB}}^{\text{H}}$, the angular distribution offers three additional null tests, one of which is the combined forward-backward asymmetry of the lepton and hadron system. As discussed in detail in Ref.~\cite{Golz:2022alh}, $F_L$ and the three forward-backward asymmetries are already sufficient to qualitatively disentangle NP effects from $C_7,\,C_7^\prime, C_9^\prime, C_{10}$ and $C_{10}^\prime$.

\section{Conclusions}\label{sec:concl}
\noindent We presented null tests in angular observables of three- and four-body decays of charmed baryons and investigated their NP sensitivity. Along with the fraction of longitudinally polarized dimuons $F_L$ and the hadron-side forward-backward asymmetry $A_{\text{FB}}^{\text{H}}$, dipole operators $O_{7}^{(\prime)}$ can be probed down to values as small as $\sim0.01$ and right-handed vector couplings $C_{9}^{\prime}$ down to $\mathcal{O}(0.1)$, whereas the null tests probe axial vector couplings $C_{10}^{(\prime)}$, which are zero in the standard model, with a significance of few$\,\times\,0.01$.
This null test strategy exploits synergies between three- and four-body modes and complements null test searches in rare charm meson decays. Contributions to different Wilson coefficients are disentangled and thus the issue of dominating resonance effects in rare charm decays is overcome. The presented strategy therefore constitutes an excellent road towards a future global fit, crucial for BSM tests with flavor changing neutral currents in the up-type flavor sector.

\acknowledgments
\noindent MG would like to thank the organizers for the wonderful web-format of the conference. This work is supported by the \textit{Studienstiftung des Deutschen Volkes} (MG) and in part by the \textit{Bundesministerium f\"ur Bildung und Forschung} (BMBF) under project number
 05H21PECL2 (GH).

\end{document}